
\hoffset=.25in
\vbadness=10000
%
%

\font\bf=cmbx10 scaled 1200

\font\it=cmti10 scaled 1200
\font\sl=cmsl10 scaled 1200

\font\tenrm=cmr10 scaled 1200
\font\sevenrm=cmr9
\font\fiverm=cmr7
\font\teni=cmmi10 scaled 1200
\font\seveni=cmmi9
\font\fivei=cmmi7
\font\tensy=cmsy10 scaled 1200
\font\sevensy=cmsy9
\font\fivesy=cmsy7

\font\tenbf=cmbx10 scaled 1200
\font\sevenbf=cmbx7 scaled 1200
\font\fivebf=cmbx5 scaled 1200
\font\tensl=cmsl10 scaled 1200
\font\tentt=cmtt10 scaled 1200
\font\tenit=cmti10 scaled 1200
\catcode`\@=11
\textfont0=\tenrm \scriptfont0=\sevenrm \scriptscriptfont0=\fiverm
\def\rm{\fam\z@\tenrm}
\textfont1=\teni \scriptfont1=\seveni \scriptscriptfont1=\fivei
\def\mit{\fam\@ne} \def\oldstyle{\fam\@ne\teni}
\textfont2=\tensy \scriptfont2=\sevensy \scriptscriptfont2=\fivesy
\def\cal{\fam\tw@}
\textfont3=\tenex \scriptfont3=\tenex \scriptscriptfont3=\tenex
\newfam\itfam \def\it{\fam\itfam\tenit} 
\textfont\itfam=\tenit
\newfam\slfam \def\sl{\fam\slfam\tensl} 
\textfont\slfam=\tensl
\newfam\bffam \def\bf{\fam\bffam\tenbf} 
\textfont\bffam=\tenbf \scriptfont\bffam=\sevenbf
\scriptscriptfont\bffam=\fivebf
\newfam\ttfam  
\textfont\ttfam=\tentt
\catcode`\@=12
%
%
\rm
\hfuzz=10pt \overfullrule=0pt
\vsize 8.75in
\hsize 6in
\baselineskip=14pt
\def\doublespace{\baselineskip=28pt}

\parindent 20pt \parskip 6pt
\def\blankline{\par\vskip \baselineskip}
\def\lbbrack{{\lbrack\!\lbrack}}
\def\rbbrack{{\rbrack\!\rbrack}}
\doublespace
\parindent=1cm
\raggedbottom
\centerline{\bf Quantum Dynamics of Lorentzian Spacetime Foam}
\blankline
\centerline{Ian H.~Redmount\footnote{*}{Permanent address:  Department of
Science and Mathematics, Parks College of Saint Louis University, Cahokia,
Illinois~~~62206.}}
\centerline{\sl Department of Physics, University of Wisconsin--Milwaukee}
\centerline{\sl Milwaukee, Wisconsin~~~53201}
\centerline{and}
\centerline{Wai-Mo Suen}
\centerline{\sl McDonnell Center for the Space Sciences}
\centerline{\sl Department of Physics, Washington University}
\centerline{\sl St.~Louis, Missouri~~~63130--4899}
\blankline
\centerline{\bf ABSTRACT}\nobreak
A simple spacetime wormhole, which evolves classically from zero throat
radius to a maximum value and recontracts, can be regarded as one possible
mode of fluctuation in the microscopic ``spacetime foam'' first suggested
by Wheeler.  The dynamics of a particularly simple version of such a wormhole
can be reduced to that of a single quantity, its throat radius; this
wormhole thus provides a ``minisuperspace model'' for a structure
in Lorentzian-signature foam.  The classical equation of motion for the
wormhole throat is obtained
from the Einstein field equations and a suitable equation of state for the
matter at the throat.  Analysis of the quantum behavior of the hole then
proceeds from an action corresponding to that equation of motion.  The action
obtained simply by calculating the scalar curvature of the hole spacetime
yields a model with features like those of the relativistic free particle.
In particular the Hamiltonian is nonlocal, and for the wormhole cannot even
be given as a differential operator in closed form.  Nonetheless the general
solution of the Schr\"odinger equation for wormhole wave functions,
i.e., the wave-function propagator, can be expressed as a path integral.
Too complicated to perform exactly, this can yet be evaluated via a WKB
approximation.  The result indicates that the wormhole, classically stable,
is quantum-mechanically unstable:  A Feynman-Kac decomposition of the
WKB propagator yields no spectrum of bound states.  Though an initially
localized wormhole wave function may oscillate for many classical
expansion/recontraction periods, it must eventually leak to large radius
values.  The possibility of such a mode unstable against growth, combined with
the observed absence of macroscopic wormholes, suggests that stability
considerations may place constraints on the nature or even the existence
of Planck-scale foamlike structure.\par
\vfill\eject
\parskip=14pt
\centerline{\bf I.~~INTRODUCTION}\par\nobreak
A quantum description of gravitation is one of the most eagerly sought goals
of present-day physics.  Approaches to such a description may be loosely
classified as ``grand schemes'' and ``small schemes.''  The former are
attempts at a comprehensive quantum theory of gravitation, such as
supergravity or superstring theory, canonical quantization of general
relativity with or without new variables, {\it et~cetera.}  The latter
are attempts to analyze the quantum-gravitational physics of particular,
simple systems, in hopes of understanding features expected to emerge from
a complete, general theory.  These small schemes include the extensive
``minisuperspace'' quantum-cosmology program~[1], the numerous calculations
of the effects of Planck-scale spacetime wormholes and ``baby universes''
on the constants of macroscopic physics~[2], and the present work, in which
we seek to probe the dynamics of such a Planck-scale ``foam'' structure
itself.  An outline of this investigation has been published previously~[3];
here we present the calculations and results in detail.

The concept of ``spacetime foam,'' first suggested by Wheeler~[4] some
35 years ago, has become standard lore in our quantum picture
of gravitation:  On scales of the Planck length, quantum fluctuations
of the spacetime geometry become so dominant that spacetime takes on
all manner of nontrivial topological structure, such as wormholes and
handles; smooth, simply connected spacetime only emerges as a classical
limit on larger scales.  Yet in the intervening decades our understanding
of quantum gravity still has not advanced sufficiently to prove or disprove
this famous conjecture.

The difficulties inherent in such a proof, or in any detailed treatment
of spacetime foam, are well known.  Topological change in a Lorentzian
manifold necessarily entails the complications of singularities or degeneracy
of the geometry, or of closed timelike curves~[5].  To circumvent these
problems one may consider Euclidean manifolds instead~[1].  Euclidean quantum
gravity is physically distinct from Lorentzian, unlike ordinary quantum
field theory in which the two formulations are equivalent via analytic
continuation.  Indeed on a quantum level it might be that spacetime is
fundamentally Euclidean, with familiar Lorentzian spacetime only a classical
limit attained far from the Planck regime.  But Euclidean quantum gravity has
its own formidable problems~[6], including the failure of the Euclidean action
to be positive definite or bounded below, the problem of interpretation, and
the matter of recovering Lorentzian spacetime.

Both versions of spacetime foam continue to be subjects of great interest.
Much recent work on Euclidean foam has explored its possible role in
determining the fundamental constants of nature~[2].  The suggestion that
Lorentzian wormholes, extracted from microscopic foam and suitably enlarged,
might conceivably be used for space or time travel~[7] has inspired a great
deal of recent and ongoing work on issues of causality and consistency in
spacetime physics.  Both sets of ideas have profound significance for
fundamental physics, but both regard the spacetime foam itself as a given,
neither probing the questions of its existence or structure substantially.

This work addresses the dynamics of Lorentzian spacetime foam, and the
implications of that dynamics for its existence and nature.  We begin by
envisioning Lorentzian spacetime, filled on Planck scales with all manner
of wormholes and other structures---modes of topological
fluctuation---continually winking into existence, persisting for microscopic
time periods, and pinching off.  The moments of creation and disappearance of
these structures, i.e., the actual points of topological change, may or may
not require a Euclideanized treatment; our analysis does not attempt to deal
with these points.

Classical wormhole geometries can be used to model such a picture.  In general
such wormholes, persisting for arbitrary lengths of time (in particular,
sufficient to allow passage of matter or energy), must involve stress-energy
distributions which violate the weak energy condition---negative energy density
in some reference frames must appear somewhere in the wormholes' throats.
Whether this suffices to rule out wormholes on macroscopic scales is unknown at
present, but it is not expected to be a problem for microscopic structure,
where quantum effects could readily give rise to such stress-energy~[7,8].
Moreover it does not appear to present a barrier to the growth of wormholes
from Planck scales to large sizes, as the present work indicates.

To make the analysis tractable we employ a brutally simplified wormhole
model to represent one mode of spacetime-foam fluctuation.  A class of
wormholes can be constructed by excising the world tube of some surface from
a $3+1$-dimensional spacetime region and joining this to another such region,
with like excision, at that surface~[9].  The join, the throat of the resulting
wormhole, contains a surface layer of stress-energy specified by the Einstein
field equations expressed as junction conditions~[10].  Here we take
the external regions of such a wormhole to be flat, empty Minkowski space, the
throat a sphere of time-varying radius~[9].  Our model is thus an extreme
version of a wormhole in which the spacetime curvature in the throat is much
greater than that in the regions surrounding the mouths.  Our simplifications
reduce the geometric degrees of freedom of the spacetime to just one, the
throat radius.  Thus our analysis is a minisuperspace model of a wormhole,
analogous to the minisuperspace models used in quantum cosmology~[1].  In fact
our model is simplified even further than those:  We treat the matter
providing the stress-energy on the wormhole throat via an equation of state,
rather than as a separate dynamical field.  The quantum/gravitational dynamics
of the complete wormhole system becomes then the quantum mechanics of one
variable, the throat radius.  The result we seek is the evolution, in time
as defined in the wormhole's flat exterior regions, of a wave function for
that radius.

In the absence of a general theory of quantum gravity, there is no
prescription from first principles for quantizing a system like this
wormhole.  The existence of an external time, in terms of which the
evolution of the system is naturally to be described, makes unsuitable
the usual time-independent Wheeler-de~Witt equation~[11].  Instead we choose
to ``solve the constraint'' explicitly, i.e., to impose the Hamiltonian
constraint classically, to reduce the phase space of the system.
Combining the classical junction conditions at the wormhole
throat---consisting of the Hamiltonian constraint plus a dynamical
equation---with an equation of state for the matter on the throat yields
an equation of motion for the throat radius alone.  We then construct an
action for the dynamics in the reduced phase space, i.e., depending only
on the radius and its time derivatives, for which that equation of motion
is the Euler-Lagrange equation.  This can be done in a variety of ways,
amenable to various forms of quantization.  One such action, associated
with gravitational dynamics in a straightforward way, is obtained from
the scalar curvature of the wormhole spacetime.  The wormhole quantum
mechanics described by this action is treated in detail in this paper.
Alternative actions and their implications for wormhole dynamics are
examined in a subsequent paper~[12].

The wormhole quantum mechanics which follows from the ``curvature action''
is in some respects similar to the Newton-Wigner first-quantized description
of a relativistic free particle~[13,14].   For both the Hamiltonian is
non-polynomial in the momentum, i.e., is a nonlocal operator in the
coordinate representation.  Indeed, for the wormhole the Hamiltonian is
a rational function of ``velocity,'' which is related to the canonical
momentum via a transcendental equation which cannot be inverted in closed
form.  This unconvential form of the Hamiltonian makes it impossible to
deduce the behavior of the system merely by examining the ``potential'' term;
explicit solution for the evolution of the wave function is needed.  Moreover
the corresponding Schr\"odinger equation for wormhole wave functions cannot
be written explicitly as a differential equation.  It can, however, be
solved, at least formally:  The wave-function propagator is obtained by
using the action in a Feynman path integral.  Exact evaluation of this
integral is problematic; the action is complicated, and questions of the
appropriate measure and skeletonization (factor-ordering problems), as well as
the issue of the correct class of paths over which to integrate, arise.  But
the integral can be evaluated in a WKB approximation, as a sum of contributions
from classical paths and small fluctuations about them.  The measure and
factor-ordering problems are thus avoided.  It is crucial, though, to
include all the appropriate paths.  The analogy to the relativistic particle
suggests that spacelike as well as timelike and lightlike paths must be
included in the integral~[14,15].  Also, since negative throat radii are
undefined, a suitable boundary condition must be imposed on the wave functions,
i.e., on the propagator, at zero radius.  Zero radius acts as a reflecting
barrier, giving rise to classical paths---extrema of the action---which
``bounce'' there one or more times.  (Such a boundary condition incorporates
our neglect of any topological change at zero radius.)  The WKB version of the
propagator, then, consists of a sum of several terms, contributions of paths
with different numbers of bounces; cancellation of terms effects the boundary
condition at zero, just as for a particle in a half space.  Explicit
expressions for all the terms are obtained here.  With these the evolution
of a wave function is calculated via a convolution integral, evaluated
numerically.

The results suggest that even with a matter equation of state chosen so
that the classical evolution of the throat radius is bounded, the wormhole
is quantum-mechanically unstable.  This is not unprecedented---classically
stable black holes, for example, are unstable to Hawking evaporation.  In this
case a Feynman-Kac decomposition of the wormhole propagator yields no spectrum
of bound states.  The calculated evolution of an initially localized
wave function follows a classical trajectory for many
expansion/recollapse/bounce cycles, but the wave function must eventually
leak to arbitrarily large throat-radius values.  Its behavior thus resembles
that for a particle initially confined to a well with finite, though perhaps
high and wide, walls, or an $\alpha$~particle in a heavy nucleus.  This simple
wormhole geometry appears to represent a mode of spacetime-foam structure
unstable against growth to macroscopic size.

Though subject to the limitations of our model, viz., the simplifications
and approximations we have applied, and the uncertainties associated with our
quantization scheme, such a conclusion is potentially very significant.  It
indicates that Planck-scale gravitational physics might be constrained by
observations on much larger scales:  The apparent absence of any topological
structure to spacetime on scales accessible to observations, ranging from
particle physics to astronomy, could imply either that not all modes of
topological fluctuation are possible or that microscopic (Lorentzian)
spacetime foam does not exist altogether.

The wormhole model and its associated classical and quantum dynamics are
presented in Sec.~II below.  Evaluation of the wormhole propagator in the
WKB approximation is shown in detail in Sec.~III.  The implications of
the result for wormhole wave-function evolution are discussed in Sec.~IV.
Conclusions and caveats follow in Sec.~V.  Units with $G=c=\hbar=1$ are
used throughout.  Sign conventions and general notation follow those of
Misner, Thorne, and Wheeler~[16].
\blankline
\centerline{\bf II.~~MECHANICS OF THE WORMHOLE MODEL}\par\nobreak
\centerline{\bf A.~Classical dynamics}\par\nobreak
A spherically symmetric ``Minkowski wormhole''~[9] provides a very simple
model of a mode of topological fluctuation in Lorentzian-signature spacetime
foam.  The classical geometry of such a wormhole is constructed by removing
a ball of time-dependent radius $r=R(t)$ from two Minkowski-space regions
and identifying the two boundary world-tubes, making them the throat of the
wormhole.  The Einstein field equations are satisfied trivially in the
flat, empty exterior regions.  At the throat those equations take the form
of the junction conditions~[10]
$$S^i_j={1\over8\pi}\,\lbbrack K^i_j-\delta^i_j\,K^m_m\rbbrack\ ,
\eqno(2.1)$$
where $S^i_j$ is the stress-energy tensor of a surface layer on the throat,
and the right-hand side is the jump discontinuity in the throat extrinsic
curvature~$K^i_j$, minus its trace~$K^m_m$.  Here the nontrivial components
of these equations are
$$\eqalignno{S_{\tau\tau}&=-{1\over2\pi R}\,{1\over(1-\dot{R}^2)^{1/2}}
&(2.2{\rm a})\cr
\noalign{\hbox{and}}
S_{\theta\theta}&={1\over4\pi}\left({R\over(1-\dot{R}^2)^{1/2}}+
{R^2\ddot{R}\over(1-\dot{R}^2)^{3/2}}\right)\ ,&(2.2{\rm b})\cr}$$
overdots denoting derivatives with respect to Minkowski-coordinate
time~$t$ (in a frame ``centered on the hole,'' i.e., in which spherical
symmetry is maintained).  The throat coordinates here are proper
time~$\tau$ and polar angle~$\theta$; proper and coordinate time are
related via $d\tau=(1-\dot{R}^2)^{1/2}dt$.  Once a surface-layer equation
of state, relating surface density $\sigma=S_{\tau\tau}$ and pressure
$p=S_{\theta\theta}/R^2$, is specified, Eqs.~(2.2) give the classical
equation of motion for the wormhole.  As is to be expected~[7,8], Eq.~(2.2a)
indicates that negative energy densities appear.

For simplicity we specify the equation of state as a property of the model,
rather than derive it from some more fundamental description of the surface
layer.  It could be chosen to make the equation of motion extremely simple.
The choice $p=-\sigma/2$, for example, would imply $\ddot{R}=0$.  The dynamics
of such a model is trivial.  Classically the wormhole throat either sits at
fixed radius or expands or contracts linearly.  Its quantum behavior is the
same as that of a free particle; a wave function initially concentrated
about some $R$~value will disperse to infinity.  But this is not a model
suitable for describing structures in spacetime foam, fluctuating into
and out of existence.  Rather we seek an equation of state such that
the equation of motion describes expansion from zero radius to some maximum
value and recollapse.  The simple ``dust'' equation of state, $p=0$, implies
the equation of motion $R\ddot{R}-\dot{R}^2+1=0$; the solutions of this are
sine functions, with the desired behavior.  This equation, however, is not
suited to the action construction we use below.  (It can be treated by
other methods~[12].)  Instead we choose
$$p=-\sigma/4\ ,\eqno(2.3)$$
which yields the equation of motion
$$2R\ddot{R}-\dot{R}^2+1=0\ .\eqno(2.4)$$
The corresponding classical trajectories are parabolic:
$$R_{\rm cl}(t)={1\over\alpha}\left(1-{\alpha^2(t-t_0)^2\over4}\right)\ ,
\eqno(2.5)$$
where $\alpha$ and $t_0$ are constants.  These solutions likewise behave as
desired.

Thus by restricting consideration to Minkowski-wormhole geometries,
and combining the classical initial-value (Hamiltonian) constraint~(2.2a)
with the dynamical equation~(2.2b) and the equation of state~(2.3), we
reduce the infinity of gravitational and matter degrees of freedom of the
wormhole spacetime to one, the throat radius~$R$.  This is thus an extreme
form of minisuperspace model.

The equation of motion~(2.4) can be obtained as the Euler-Lagrange equation
for extrema of the action
$$S=\int\left(R\dot{R}\ln\left|{1+\dot{R}\over1-\dot{R}}\right|-2R\right)\,dt
\ .\eqno(2.6)$$
This choice of reduced-phase-space action is motivated by the fact
that it coincides with the Einstein-Hilbert action for the wormhole spacetime.
The wormhole's scalar curvature~${\cal R}$ is nonzero only on the throat:
$$\eqalignno{{\cal R}&=2\lbbrack K^m_m\rbbrack\delta(\rho)&\cr
&={4\over R^2(1-\dot{R}^2)^{1/2}}
\left[R\dot{R}\ln\left|{1+\dot{R}\over1-\dot{R}}\right|-2R
-{\textstyle{1\over2}}{d\over dt}
\left(R^2\ln\left|{1+\dot{R}\over1-\dot{R}}\right|\right)\right]
\delta(\rho)\ ,&\cr
&&(2.7)\cr}$$
where $\rho$ is $r-R(t)$ in one exterior region and $R(t)-r$ in the other.
The corresponding gravitational action---the integral of~${\cal R}$ plus
a surface term~[17] eliminating the total-derivative term---is just Eq.~(2.6).

The Hamiltonian dynamics of the wormhole, on its reduced phase space, follows
immediately from the action.  With Lagrangian~$L$ given by the integrand in
Eq.~(2.7), the canonical momentum is
$$P={\partial L\over\partial\dot{R}}=
R\left(\ln\left|{1+\dot{R}\over1-\dot{R}}\right|+{2R\dot{R}\over1-\dot{R}^2}
\right)\ .\eqno(2.8)$$
The Hamiltonian is
$$H=P\dot{R}-L={2R\over1-\dot{R}^2}\ .\eqno(2.9)$$
The Hamiltonian cannot be expressed in terms of~$R$ and~$P$ in closed form,
since the transcendental relation~(2.9) between~$\dot{R}$ and~$P/R$ cannot be
inverted explicitly.

It should be noted that the ``energy'' corresponding to this Hamiltonian is
not to be identified with the ADM mass of the wormhole, which is zero by
construction.  To thus fix the ``energy'' would reduce the equation of motion
to first order, and the classical initial-parameter space to one dimension,
making canonical quantization impossible.  In fact the Hamiltonian~(2.9)
corresponds to the classically conserved ``mass''~$2(2\pi R^{3/2}\sigma)^2$
appropriate to the equation of state~(2.3).  [The Einstein field equations
imply the conservation law $d(4\pi R^2\sigma)/dt+pd(4\pi R^2)/dt=0$; Eq.~(2.3)
then implies that $R^{3/2}\sigma$, hence $H$, are classically conserved.]
This can take on various values for different wormhole states.  In this
respect our treatment contrasts with the analyses of H\'aj\'\i\v{c}ek
{\it et.~al.}~[18] of the quantum mechanics of a collapsing dust shell.
Those authors take the Hamiltonian to correspond to the ADM mass of the
system, which may take on different values in different states.  They treat
the classically conserved proper mass of the shell as a fixed parameter.

Of course the wormhole action is not uniquely defined by the equation of
motion specifiying its extrema.  A variety of actions, of other than
Einstein-Hilbert form, can be constructed corresponding to Eq.~(2.4)---or
to equations obtained from other equations of state.  Actions can be
obtained which avoid some of the complexities encountered with the geometric
action~(2.6), e.g., the complicated kinetic term, and the transcendental
relation~(2.8).  The construction of such actions and their implications for
wormhole quantum mechanics are examined in another paper~[12].  The rest of
this paper treats the quantum dynamics obtained from the geometric action.
\blankline
\centerline{\bf B.~Quantum dynamics}\par\nobreak
In this minisuperspace model, then, the quantum wormhole is to be
described by a wave function~$\psi(R,t)$.  With the Hamiltonian
constraint~(2.2a) imposed classically, i.e., incorporated into the
reduction of the wormhole's phase space, no Wheeler-de~Witt equation
imposes ``time independence'' of~$\psi$~[11]; instead its evolution
in wormhole-exterior time~$t$ is generated by the Hamiltonian~(2.9).

There is no hope of treating the corresponding Schr\"odinger equation
directly:  The transcendental nature of relation~(2.8) and the form of~$H$
imply that the Hamiltonian contains all powers of the momentum~$P$, or as an
operator on~$\psi$, of $\partial/\partial R$.  It is therefore a nonlocal
operator, similar to that for a relativistic particle, but rather more
unwieldy~[14].  Operator ordering in~$H$ is problematic as well.
But the {\it solution\/} of the Schr\"odinger equation can be treated.
The general solution, i.e., the time evolution of any wave function,
can be given in terms of a propagator:
$$\psi(R,t)=\int G[R,t;R_0,0]\,\psi_0(R_0)\,dR_0\ .\eqno(2.10)$$
The propagator~$G$ is given formally by a Feynman path integral
$$G[R,t;R_0,0]=\int_C e^{iS[R(t)]}\,{\cal D}[R(t)]\ ,\eqno(2.11)$$
with $C$ denoting the class of paths over which the integral is defined.

The appropriate class of paths is an important issue.  The example
of the relativistic particle~[14], and more general quantum-gravitational
considerations~[15], suggest that all paths linking the initial and
final points should be included---spacelike as well as timelike and null.
[Spacelike paths, with $|\dot{R}|>1$, correspond to ``negative energy''
contributions, in terms of the Hamiltonian~(2.9).]  Including spacelike paths
implies that $G$ need not vanish outside the ``light cone'' $|R-R_0|=t$, i.e.,
that it may admit ``acausal propagation.''

Moreover, the wormhole geometry is only defined for nonnegative~$R$, hence
only paths with $R(t)\ge0$ should be included.  This restriction can be
implemented as for a particle confined to a half space, i.e., as if
there were an infinite potential wall at~$R=0$.  This implies the
boundary condition $\psi(0,t)=0$, hence that $G[R,t;R_0,0]$ should
vanish for~$R$ or~$R_0$ zero.  (More precisely, $\psi$ must entail
no current in the $-R$~direction at $R=0$, so boundary conditions more
general than this are possible.  For simplicity we do not treat such here.)
By imposing this condition we exclude consideration of topology-changing
processes, i.e., wormhole creation or disappearance, at~$R=0$.  The treatment
of such processes might require a ``second quantized'' framework (actually
``third quantized,'' outside the restrictions of a minisuperspace model),
rather than the ``first quantized'' (actually ``second quantized'') description
used here.  Their bearing on our results is discussed further in Sec.~V.
The condition on~$\psi$ and~$G$ gives $R=0$ here the role of a reflecting
boundary for wormhole wave functions.
\blankline
\centerline{\bf III.~~EVALUATION OF THE PROPAGATOR:
WKB APPROXIMATION}\par\nobreak
Exact evaluation of the path integral~(2.11) is out of reach:  The integral,
with action~(2.6), is non-Gaussian; the operator-ordering problems involved
in canonical quantization with the Hamiltonian~(2.9) would reappear as
ambiguities in the skeletonization of the path integral; the appropriate
measure on the space of paths is unknown.  But over a suitable range of
time intervals and throat-radius values it should be possible to calculate
the propagator via a WKB approximation.  Then the path integral is taken to be
dominated by the contributions of classical trajectories---local extrema of
the action---and nearby paths.  The approximate propagator takes the form
$$G^{\rm (WKB)}[R,t;R_0,0]=\sum_{{\rm Classical}\atop{\rm Paths}}
\left({i\over2\pi}\,{\partial^2S[R_{\rm cl}]\over\partial R\partial R_0}
\right)^{1/2}\,e^{iS[R_{\rm cl}]}\ .\eqno(3.1)$$
The exponential factors arise from the classical paths~$R_{\rm cl}$ between
the initial and final values.  The prefactors come from Gaussian integrals
over small deviations from these, giving semiclassical corrections~[19].
To that order $G^{\rm (WKB)}$ is independent of the choice of path-integral
measure.  Specifying the regime in which this approximation is accurate is
problematic, since we have neither an exact solution for comparison nor even
an explicit form of the Schr\"odinger equation.  This question is considered
further in Sec.~V below.

The first step in evaluating $G^{\rm (WKB)}$, then, is finding the
appropriate classical paths.  More than one type of path always contributes.
There is exactly one parabolic trajectory of form~(2.5), and taking only
nonnegative radius values, between any initial and final points~$(R_0,0)$
and~$(R,t)$.  Specifically, this trajectory has parameters
$$\eqalignno{\alpha&={8\over t^2}\left[\left(R_0R+{t^2\over4}\right)^{1/2}
-{R_0+R\over2}\right]&(3.2{\rm a})\cr
\noalign{\hbox{and}}
t_0&={t\over2}\,{\displaystyle{\left(R_0R+{t^2\over4}\right)^{1/2}-R_0}\over
\displaystyle{\left(R_0R+{t^2\over4}\right)^{1/2}-{R_0+R\over2}}}
\ ,&(3.2{\rm b})\cr}$$
and action
$$\eqalignno{S^{(0)}=&-{3t\over2\alpha}\left[1-{\alpha^2\over12}
(t^2-3tt_0+3t_0^2)\right]\cr
&\qquad+{R^2\over2}\ln\left|{2-\alpha(t-t_0)\over2+\alpha(t-t_0)}\right|
-{R_0^2\over2}\ln\left|{2+\alpha t_0\over2-\alpha t_0}\right|\ .&(3.3)\cr}$$
Because paths are restricted to the half-space of positive radius values,
there are also classical trajectories which are piecewise of form~(2.5), but
which ``bounce'' one or more times at zero radius; extremizing the action
determines the bounce times.  Such a path which bounces just once, at
time~$t_1$, has action
$$\eqalignno{S^{(1)}=&-{\textstyle{1\over4}}[t_1(t_1+2R_0)+
(t-t_1)(t-t_1+2R)]\cr
&\qquad+{\textstyle{1\over2}}\left(R^2\ln\left|{R\over t-t_1-R}\right|
-R_0^2\ln\left|{t_1-R_0\over R_0}\right|\right)\ .&(3.4)\cr}$$
For the bounced path to contribute to the WKB propagator~(3.1), this should
be extremal with respect to variation of~$t_1$.  That condition is
$$(t-t_1)^2(t_1-R_0)-t_1^2(t-t_1-R)=0\ .\eqno(3.5)$$
Each real solution of this equation with~$t_1\in[0,t]$ corresponds to a
contributing trajectory.  Since the left-hand side of Eq.~(3.5) changes
sign between~$t_1=0$ and~$t_1=t$ there is always one such solution,
and may be three.   A classical path bouncing more than once, at
times~$t_1,\ldots,t_n$, has action
$$\eqalignno{S^{(n)}=&-{\textstyle{1\over4}}[t_1(t_1+2R_0)
+\sum_{j=2}^n(t_j-t_{j-1})^2+(t-t_n)(t-t_n+2R)]\cr
&\qquad+{\textstyle{1\over2}}\left(R^2\ln\left|{R\over t-t_n-R}\right|
-R_0^2\ln\left|{t_1-R_0\over R_0}\right|\right)\ .&(3.6)\cr}$$
That this be extremal with respect to all bounce times is equivalent to
the conditions
$$(n-1)t_1^2-(t_1-R_0)(t_n-t_1)=0\ ,\eqno(3.7{\rm a})$$
$$t_j=t_1+{j-1\over n-1}(t_n-t_1)\ ,\qquad 2\le j\le n-1\ ,\eqno(3.7{\rm b})$$
and
$$(n-1)(t-t_n)^2-(t_n-t_1)(t-t_n-R)=0\ .\eqno(3.7{\rm c})$$
Any real solution to the coupled quadratic equations~(3.7a) and~(3.7c)
with~$0\le t_1<t_n\le t$ yields an $n$-bounce path contributing
to~$G^{\rm (WKB)}$.  For a given $n$ there may be zero, two, or four such.
Conditions~(3.5) and~(3.7a--c) are equivalent to requiring that the
parameters~$\alpha$ for all the parabolic segments of a bouncing path be
equal, which is to say that the wormhole ``bounces elastically.''

For any set of initial and final coordinates~$(R_0,0)$ and~$(R,t)$ there
is a maximum number of bounces $n_{\rm max}$, beyond which there are no
appropriate solutions to Eqs.~(3.7a) and~(3.7c) and no contributions
to~$G^{\rm (WKB)}$.  For example, in the case $R_0=R$, the four solutions
of Eqs.~(3.7a) and~(3.7c) are
$$\eqalign{t_1^{(a)}&=\left(t-\sqrt{t^2-4Rnt}\right)/(2n)\cr
t_n^{(a)}&=\left[(2n-1)t-\sqrt{t^2-4Rnt}\right]/(2n)\ ,\cr}\eqno(3.8{\rm a})$$
\vskip 7pt
$$\eqalign{t_1^{(b)}&=\left(t+\sqrt{t^2-4Rnt}\right)/(2n)\cr
t_n^{(b)}&=\left[(2n-1)t+\sqrt{t^2-4Rnt}\right]/(2n)\ ,\cr}\eqno(3.8{\rm b})$$
\vskip 7pt
$$\eqalign{t_1^{(c)}&=\left(t+2R+\sqrt{t^2-4Rnt+4R^2}\right)/[2(n+1)]\cr
t_n^{(c)}&=\left[(2n+1)t-2R-\sqrt{t^2-4Rnt+4R^2}\right]/[2(n+1)]\ ,\cr}
\eqno(3.8{\rm c})$$
\noindent and
$$\eqalign{t_1^{(d)}&=\left(t+2R-\sqrt{t^2-4Rnt+4R^2}\right)/[2(n+1)]\cr
t_n^{(d)}&=\left[(2n+1)t-2R+\sqrt{t^2-4Rnt+4R^2}\right]/[2(n+1)]\ .\cr}
\eqno(3.8{\rm d})$$
For~$n$ between~2 and~$t/(4R)$, if any, all four solutions are real and
in the desired interval; four $n$-bounce paths contribute to~$G^{\rm (WKB)}$.
For any $n$ between~$t/(4R)$ and~$(t^2+4R^2)/(4Rt)$ the $(a)$ and $(b)$
solutions are complex, and there are no corresponding classical paths, but
the $(c)$ and $(d)$ solutions characterize two $n$-bounce paths contributing
to the propagator.  For $n$ greater than $(t^2+4R^2)/(4Rt)$ there are no
$n$-bounce classical paths.  (The values $R_0=R=0$, for any $t$, are a
limiting case, in which four $n$-bounce paths contribute for all $n$ greater
than 1.)  It must be emphasized that these particular bounds for~$n$, and
the classification of the solutions $(a)$, $(b)$, $(c)$, and $(d)$, cannot
be applied for arbitrary unequal~$R_0$ and~$R$, but there are always some
bounds on the $n$~values of contributing paths.  This is because between
bounces the paths are of form~(2.5), for which there is a fixed ratio between
maximum radius and the interval between zero crossings.  For any $t$~value,
then, there is time for only so many bounces high enough to reach a given
$R_0$~or $R$~value.

The evaluation of~$G^{\rm (WKB)}$ proceeds from these results.
Expression~(3.1) eventually takes the form
$$G^{\rm (WKB)}[R,t;R_0,0]=\sum_{n=0}^{n_{\rm max}(R_0,R,t)}\sum_k
P_k^{(n)}\left({P_k^{(n)}F_k^{(n)}\over4\pi i}\right)^{1/2}
\exp\left(iS_k^{(n)}\right)\ ,\eqno(3.9)$$
with $n$ the number of bounces, as above, and $k$ labelling the
$n$-bounce paths, the range of~$k$ depending of course on~$n$.  The
classical actions~$S_k^{(n)}$ are given by Eqs.~(3.3), (3.4), and~(3.6).
The functional determinants~$F_k^{(n)}$ are:
$$F^{(0)}={32/(\alpha t)\over4+\alpha^2t_0(t-t_0)}\ ,\eqno(3.10{\rm a})$$
with $\alpha$ and $t_0$ from Eqs.~(3.2);
$$F^{(1)}=-{t_1^4/(t_1-R_0)^2\over t^2+2R_0t-6tt_1+6t_1^2+2(R-R_0)t_1}\ ,
\eqno(3.10{\rm b})$$
with $t_1$ from Eq.~(3.5);
$$F^{(n)}={t_1^2(t_n-t_1)/(t_1-R_0)\over{\displaystyle{
t[(4n^2-2n-1)t_1-(2n-1)t_n-2(n-1)R_0]+R[t_n-(2n-1)t_1]\atop
+2n(t_1^2+t_n^2)-4n^2t_1t_n+R_0[(2n-1)t_n-t_1]}}}\ ,\eqno(3.10{\rm c})$$
for $n\ge2$, with $t_1$ and $t_n$ from Eqs.~(3.7a) and~(3.7c).  The phase
factors are $P_k^{(n)}=+1$ if $F_k^{(n)}$ has the same sign as $F^{(0)}$ and
$P_k^{(n)}=-1$ if it has the opposite sign.  This choice of phases ensures
the boundary conditions $G^{\rm (WKB)}[0,t;R_0,0]=0$ and
$G^{\rm (WKB)}[R,t;0,0]=0$, hence the condition $\psi(0,t)=0$ for any
wormhole wave function, via the cancellation of contributions from paths
differing by one bounce.  For example, as $R$ approaches zero, the contribution
of the direct (zero-bounce) classical path becomes equal in magnitude and
opposite in sign to that of a one-bounce path\footnote{${}^1$}{The absolute
value of the argument of the logarithm is taken in the action~(2.6) and
subsequent expressions to ensure that this cancellation still occurs when
the direct path is just timelike and the one-bounce path just spacelike, i.e.,
for~$R_0\approx t$ and $R$ approaching zero.} with bounce time approaching~$t$.
This phase prescription accords with that used, for example, for the simple
problem of a particle confined to a half space~[20].

Form~(3.9) and the accompanying expressions show that even the approximate
propagator for this simplified system is a complicated quantity.  Some of its
features are illustrated in Fig.~1.  The propagator has singularities on
the outgoing and ingoing or ``bounced'' light cones $R=R_0+t$, $R=R_0-t$,
and $R=t-R_0$, respectively.  (It has other singular points as well, described
below.)  As expected, given the inclusion of spacelike trajectories in the
path integral as discussed in Sec.~II, the propagator is nonvanishing outside
the light cone.  Thus it admits acausal, i.e., superluminal, propagation.
Overlying these gross features the propagator exhibits a great deal of
high-frequency structure.
\blankline
\centerline{\bf IV.~~QUANTUM EVOLUTION OF THE WORMHOLE}\par\nobreak
In principle the propagator contains a complete description of the
quantum dynamics of the wormhole.  Analysis of the approximate
form~$G^{\rm (WKB)}$, carried out in several stages, indicates that
the wormhole is quantum-mechanically unstable against growth to large size.

Straightforward numerical evaluation of the propagation integral~(2.10),
using propagator~$G^{\rm (WKB)}$ as given by Eq.~(3.9) and a simple initial
wave function, reveals behavior of the wormhole over short and intermediate
time scales.  Results of such calculations are displayed in Fig.~2.  For this
example the initial wave function is a localized ``wave packet,''
$\psi_0=(2/\pi)^{1/4}\exp[-(R_0-10)^2]$, all quantities in Planck units.
For the purpose of numerical integration this is taken to be zero outside
the interval~$6\le R_0\le14$.  Although all terms in Eq.~(3.9) have been
given analytically, the numerical integration is nontrivial due to the
existence of singular points in~$G^{\rm (WKB)}$.  The poles in the propagator
on the outgoing, incoming, and bounced light cones are treated numerically
by replacing the divergent term in $G^{\rm (WKB)}$, in a small interval
(denoted~$\Delta$) in~$R_0$ about the light cone, by its average over that
interval.  This is determined from an expansion in~$R_0$ of the appropriate
term about the light cone.  In particular we obtain
$$\displaylines{\quad\int_{R-t-\Delta}^{R-t+\Delta}\left({F^{(0)}\over
4\pi i}\right)^{1/2}\exp\left(iS^{(0)}\right)\,dR_0=\left({t(2R-t)\over
4\pi i}\right)^{1/2}{4i\over t(2R-t)}\hfill\cr
\hfill{}\times\exp\Biggl\{{i\over2}\left[t(2R-t)\ln\left({t(2R-t)\over R\Delta}
\right)+(R-t)^2\ln\left({R-t\over R}\right)\right]\quad\cr
\hfill{}-{3i\over4}t(2R-t)\Biggr\}+O(\Delta^2\ln\Delta)\quad\cr
\hfill(4.1{\rm a})\cr}$$
on the outgoing light cone,
$$\displaylines{\quad\int_{R+t-\Delta}^{R+t+\Delta}\left({F^{(0)}\over
4\pi i}\right)^{1/2}\exp\left(iS^{(0)}\right)\,dR_0=\left({t(2R+t)\over
4\pi i}\right)^{1/2}{4i\over t(2R+t)}\hfill\cr
\hfill{}\times\exp\Biggl\{{i\over2}\left[t(2R+t)\ln\left({t(2R+t)\over R\Delta}
\right)+(R+t)^2\ln\left({R\over R+t}\right)\right]\quad\cr
\hfill{}-{3i\over4}t(2R+t)\Biggr\}+O(\Delta^2\ln\Delta)\quad\cr
\hfill(4.1{\rm b})\cr}$$
on the ingoing light cone, and
$$\displaylines{\int_{t-R-\Delta}^{t-R+\Delta}P^{(1)}\left(
{P^{(1)}F^{(1)}\over4\pi i}\right)^{1/2}\exp\left(iS^{(1)}\right)\,dR_0
=-\left({R^2+(t-R)^2\over4\pi i}\right)^{1/2}{4i\over R^2+(t-R)^2}\hfill\cr
\hfill{}\times\exp\Biggl\{{i\over2}\left[R^2\ln\left({R^2+(t-R)^2\over R\Delta}
\right)-(t-R)^2\ln\left({(t-R)\Delta\over R^2+(t-R)^2}\right)\right]\cr
\hfill{}-{3i\over4}[R^2+(t-R)^2]\Biggr\}+O(\Delta^2\ln\Delta)\cr
\hfill(4.1{\rm c})\cr}$$
on the bounced light cone (where only one single-bounce, and no
multiple-bounce,
classical path contributes).  In the indicated intervals, then, these singular
terms in the propagator are replaced by~$1/(2\Delta)$ times the right-hand
sides of these expressions.  We have checked that although the integrals~(4.1)
depend on~$\Delta$, the final integrals~(2.10) are independent of~$\Delta$
over a range of values from~$10^{-4}$ to~$10^{-1}$ Planck lengths.  The WKB
propagator has additional singularities, caustics, where new bouncing
paths appear, i.e., where Eq.~(3.5) or Eqs.~(3.7a and~c) have coincident
roots.  These are avoided by evaluating Eq.~(2.10) at $t$~values which are
transcendental to machine accuracy; rational values of~$R$ and~$R_0$ are taken
so the $t$~values at which caustics would occur are algebraic or rational.
Figure~2, then, shows the resulting wave function~$\psi(R,t)$.  Its most
prominent feature is that the wave packet follows an essentially classical
bouncing trajectory.  Some rapid spreading of the packet appears early on,
but this behavior does not seem to recur at later times.  Numerical
difficulties hinder the reliable evaluation of~$\psi$ at $t$~values larger
than those shown:  Accurate numerical integration becomes problematic as the
propagator develops ever higher-frequency behavior, and the problem of
accurately solving Eqs.~(3.5) and~(3.7a and~c) with large coefficients makes
even calculating the propagator difficult.  To determine the long-term behavior
and stability of the wormhole a different sort of analysis is required.

The late-time behavior of the propagator can be examined analytically in
a special, but very useful, case.  For arbitrary $R_0$, $R$, and~$t$ the
explicit form of~$G^{\rm (WKB)}$ is intractable---the general expressions
for the bounce times~$t_1$ and~$t_n$ appearing in the~$S^{(n)}$
and~$F^{(n)}$ are too unwieldy even for computer manipulation!  But
for~$R_0=R$ the solutions~(3.8) yield manageable results, and nearby
$R_0$~and $R$~values can be handled perturbatively.  The terms in the
sum~(3.9) are then labeled $a$, $b$, $c$, and~$d$, corresponding
to those solutions.  In the limit~$t\gg R,R_0$, with~$R=R_0+\varepsilon$
and~$|\varepsilon|\ll R_0$, the functional determinants and classical actions
in those terms are:
$$F_a^{(n)}=F_b^{(n)}=-{1\over n}\left[1+{4nR_0\over t}
+{16n^2R_0^2\over t^2}+O\left({n^3R_0^3\over t^3}\right)\right]
+O\left({\varepsilon\over R_0}\right)\ ,\eqno(4.2{\rm a})$$
$$S_a^{(n)}=-{t^2\over4n}-\left[{t\over2nR_0}+\ln\left({t\over nR_0}\right)
-1-{3nR_0\over t}+O\left({n^2R_0^2\over t^2}\right)\right]\,R_0\varepsilon
+O\left({t\over nR_0}\varepsilon^2\right)\ ,\eqno(4.2{\rm b})$$
and
$$S_b^{(n)}=-{t^2\over4n}+\left[{t\over2nR_0}+\ln\left({t\over nR_0}\right)
-1-{3nR_0\over t}+O\left({n^2R_0^2\over t^2}\right)\right]\,R_0\varepsilon
+O\left({t\over nR_0}\varepsilon^2\right)\ ,\eqno(4.2{\rm c})$$
for $n$ less than $t/(4R_0)$; and
$$F_c^{(n)}={1\over n+1}\left[1+{4(n+1)R_0\over t}+{2(n+1)(8n+3)R_0^2\over t^2}
+O\left({n^3R_0^3\over t^3}\right)\right]+O\left({\varepsilon\over R_0}\right)
\ ,\eqno(4.2{\rm d})$$
$$\displaylines{S_c^{(n)}=-{t^2\over4(n+1)}-{R_0t\over n+1}
-R_0^2\ln\left({t\over(n+1)R_0}\right)+{R_0^2\over2}{n-1\over n+1}
+{2nR_0^3\over t}+O\left({n^2R_0^4\over t^2}\right)\hfill\cr
\hfill-\left[{t\over2(n+1)R_0}+\ln\left({t\over(n+1)R_0}\right)-{n\over n+1}
-{3nR_0\over t}+O\left({n^2R_0^2\over t^2}\right)\right]\,R_0\varepsilon\cr
\hfill+O\left({t\over(n+1)R_0}\varepsilon^2\right)\ ,\hfill(4.2{\rm e})\cr}$$
$$F_d^{(n)}={1\over n-1}\left[1+{4(n-1)R_0\over t}+{2(n-1)(8n-3)R_0^2\over t^2}
+O\left({n^3R_0^3\over t^3}\right)\right]+O\left({\varepsilon\over R_0}\right)
\ ,\eqno(4.2{\rm f})$$
$$\displaylines{S_d^{(n)}=-{t^2\over4(n-1)}+{R_0t\over n-1}
+R_0^2\ln\left({t\over(n-1)R_0}\right)-{R_0^2\over2}{n+1\over n-1}
-{2nR_0^3\over t}+O\left({n^2R_0^4\over t^2}\right)\hfill\cr
\hfill+\left[{t\over2(n-1)R_0}+\ln\left({t\over(n+1)R_0}\right)-{n\over n-1}
-{3nR_0\over t}+O\left({n^2R_0^2\over t^2}\right)\right]\,R_0\varepsilon\cr
\hfill+O\left({t\over(n+1)R_0}\varepsilon^2\right)\ ,\hfill(4.2{\rm g})\cr}$$
for $n$ less than $(t^2+4R_0^2)/(4R_0t)$.  The $n=0$~term is a $c$~term, given
by Eqs.~(4.2d) and~(4.2e), and the $n=1$ terms are $a$, $b$, and $c$~terms
given
by Eqs.~(4.2a--e).  The order-$\varepsilon$ terms in the actions are obtained
from the relation $\partial S_{\rm cl}/\partial R=P(t)$ and Eq.~(2.8) for the
momentum~$P$, evaluated on the appropriate classical path.  Terms of
order~$\varepsilon$ and higher in the functional determinants give rise to
subdominant contributions and are not needed here.

The above expansions in inverse powers of~$t$ do not converge for
$n=t/(4R_0)$.  Indeed, the $a$, $b$, and $c$~terms in the propagator
are singular for these parameter values, at which the roots~(3.8a--c)
coincide.  But the contribution of these singular terms to the evolution
of a wave function can still be estimated.  At fixed $R_0$, such a singularity
or ``caustic point'' occurs at intervals in~$t$ of~$4R_0$, the period
of a classical bouncing trajectory with maximum radius value~$R_0$.
The singular contribution is thus associated with the classically
bouncing peak seen in Fig.~2.  Unlike the similar singularity in a
harmonic-oscillator propagator~[21], however, this singularity
in~$G^{\rm (WKB)}$ does not correspond to a $\delta$~function.  Rather,
in the immediate vicinity of the caustic, i.e., for~$\varepsilon$ small
compared to~$t^{-2}$ (in Planck units), the $c$~term has functional determinant
$$F_c^{(n)}\sim {-8R_0\over3t^{1/3}\varepsilon^{2/3}}+\cdots\eqno(4.3{\rm a})$$
and action
$$S_c^{(n)}\sim-R_0t-{R_0t^{1/3}\over2}\,\varepsilon^{2/3}+\cdots\ .
\eqno(4.3{\rm b})$$
The contribution to a wave function from this vicinity is therefore of
order~$t^{-3/2}$, not unity, as a $\delta$~function would yield.  However,
the form of the action~(4.3b) suggests that a significant contribution from
this term may arise from a region of width~$t^{-1/2}$ in~$\varepsilon$.  Such
a contribution might be of order unity, as Fig.~2 suggests, but the
expansions used in Eqs.~(4.3) do not suffice to calculate it precisely.
The $a$~and~$b$ terms do not in fact exist for nonzero~$\varepsilon$ about
the caustic point; they give no contribution here.  Another set of caustic
points occurs where the $c$~and~$d$ roots coincide.  Here again the
$a$~and~$b$ terms do not contribute, those roots being complex, while the
$c$~and~$d$ terms are similar in form to those given by Eqs.~(4.3).

Hence it is easiest to examine the behavior of~$G^{\rm (WKB)}$ for parameter
values such that no caustics occur at integer $n$~values.  Then the
expansions~(4.2) give approximate values for all terms.  The actions vary
slowest with~$\varepsilon$, i.e., with~$R_0$, for the largest-$n$ terms, so
it is these which determine the width of the region in integral~(2.10) which
gives a significant contribution to a wave function.  That width corresponds
roughly to one oscillation of the slowest-varying terms.  For example, the
integral of~$G^{\rm (WKB)}[R,t;R-\varepsilon,0]$, with terms given by
Eqs.~(4.2), over the interval $\varepsilon\in[-\pi/R,+\pi/R]$, is shown in
Fig.~3 for fixed~$R$ and various $t$~values.  These results suggest that
a wave function at fixed radius value continues to oscillate even for quite
large~$t$ values, as might be expected from a sum of terms with rapidly
varying phases.  Thus they still give no clear picture of the very-long-term
behavior of the wormhole.

But telling features of that behavior can be extracted from the propagator
as approximated by Eqs.~(4.2).  Did the Hamiltonian~(2.9) describe
a bound system, its ground-state energy (or, for a free system, the bottom
of its continuous spectrum) would be given by the Feynman-Kac~[22] limit
$$E_0=-\lim_{\tau\to+\infty}{1\over\tau}\,\ln G[R,-i\tau;R,0]\eqno(4.4)$$
of the propagator in imaginary time.  This can be evaluated via Eqs.~(4.2).
The propagator is still a sum of terms with rapidly varying phases with
respect to increasing~$\tau$, owing to the $t^2$~terms in the actions.
Consequently it oscillates forever; the limit on the right side of Eq.~(4.4)
does not exist.  This is illustrated in Fig.~4.  This means that the quantum
wormhole possesses no spectrum of bound states.  The expectation value of~$R$
cannot be confined in the late-time limit.  As for a system with an inverted
potential diverging to negative infinity, or a metastable system such as a
particle confined by finite walls, the wormhole wave function must eventually
run or ``leak'' to arbitrarily large radius values---the wormhole is unstable
against eventual growth to large size.  The previous results suggest that it
is the slow leaking behavior which characterizes these wormholes.  They appear
reminiscent of $\alpha$~particles in a nucleus, oscillating perhaps millions
of times before escaping to infinity.
\blankline
\centerline{\bf V.~~IMPLICATIONS AND LIMITATIONS}\par\nobreak
Spherically symmetric Minkowksi wormholes~[9] provide a simple model of a
mode of topological fluctuation in Lorentzian spacetime foam, a mode
apparently unstable against growth to macroscopic size.  The
quantum-gravitational dynamics of these wormholes is reduced to the
quantum mechanics of a single variable, the throat radius, by describing
the matter at the wormhole throat with a suitable equation of state and
imposing the Hamiltonian constraint at the classical level to reduce the
phase space of the system.  The corresponding reduced action is used in
a Feynman path integral to obtain the propagator for wormhole wave functions;
this is evaluated in the WKB approximation.  The result indicates that
these wormholes have no bound quantum states.  Though their throat radii
are classically bounded, i.e., they are classically stable, they
will nonetheless grow to large size by quantum ``diffusion.''

Many systems exhibit similar behavior.  For a particle with the familiar
quadratic kinetic term in the action, the form of the potential
determines whether such diffusion or spreading occurs:  A potential well
with walls or barriers which fall off at large distances will allow a
classically bound particle to leak out via quantum tunneling (as in the
case of $\alpha$~decay), while one which increases monotonically with
distance will not.  For these wormholes, with more complicated action~(2.6),
so simple an analysis is not possible.  The more involved examination of
the wormholes' quantum dynamics described here is needed to see that spreading
of the wave function to large radii will take place.

Our result suggests that dynamics, and stability considerations in particular,
may be of great importance in understanding the quantum nature of gravitation.
A definitive demonstration of the existence of an unstable mode of fluctuation
in spacetime foam would have profound implications.  Since a macroscopic
structure of wormholes is not observed, i.e., spacetime appears to be smooth
and topologically trivial on all scales accessible to laboratory physics,
it would imply the existence of a mechanism for suppressing such a mode,
or even the absence of (Lorentzian-signature) spacetime foam altogether.
Of course the present work is far from such a definitve demonstration; it
serves to point up lines along which these matters should be studied further.

The most fundamental limitation of our calculation is the restriction of
the gravitational degrees of freedom to those of the spherically symmetric
Minkowski wormhole, i.e., the use of a ``minisuperspace model'' for
topological structure.  In fact our model is even more restricted than
the usual minisuperspace models~[1], since the matter in the hole is treated
not as a dynamical field but by the use of an equation of state.  Moreover we
use the particular equation of state~(2.3), to simplify the calculations;
other possible choices are considered in Ref.~12.

We analyze our constrained model by quantization in the reduced phase space,
as described above.  In the absence of a general framework for quantum-gravity
calculations, this method seems best suited to the problem.  It does differ
markedly, though, from the Wheeler-de~Witt approach~[11].

Furthermore, we use the particular reduced action~(2.6).  Other forms
corresponding to the classical equation~(2.4) are possible; the effect
of this choice on the results will be examined elsewhere~[12].

Moreover our choice of action implies a choice of Hamiltonian fundamentally
different from that used in similar calculations~[18], corresponding here
to a classically conserved ``proper mass'' rather than the ADM mass of the
wormhole.  This choice raises some interesting questions.  For example, since
the Hamiltonian~(2.9) is time independent, its expectation value is
conserved.  Its form appears to suggest, then, that the expectation value
of the throat radius~$R$ should remain bounded.  It might be expected, however,
that any initially localized wave function must contain both positive-
and negative-``energy'' ($|\dot{R}|>1$) components; this is a well-known
feature of, e.g., the quantum mechanics of a relativistic particle~[13,14].
The interplay of these components could account for the eventual spreading
to large radius values.  More detailed calculations of the wormholes' quantum
dynamics should clarify this.

We employ the WKB approximation to evaluate the wormhole propagator.  WKB
calculations of quantum instabilities in classically stable systems, such
as tunnelling, are well known.  Here it is more difficult to be precise
about the accuracy of the approximation.  It is certainly to be expected
to be valid in the late-time limit~$t\gg R,R_0$, in which the instability is
manifest.  But the accuracy of the numerical evolution of a wave function,
as shown in Fig.~2, is harder to establish.  Lacking any exact solution for
comparison, we have tested the accuracy of the calculation via the composition
relation
$$G^{\rm (WKB)}[R,t;R_0,0]=\int G^{\rm (WKB)}[R,t;R_1,t_1]
G^{\rm (WKB)}[R_1,t_1;R_0,0]\,dR_1\ ,\eqno(5.1)$$
with $0<t_1<t$.  It can be shown analytically that this should hold if the
WKB approximation is strictly valid at the intermediate time~$t_1$.  We found,
however, that for some $t_1$~values relation~(5.1) is not well satisfied.
This may indicate inaccuracy of the WKB approximation, or may be due to
numerical difficulties associated with the integration, given the rapidly
varying phase of the propagator.

A final limitation, of fundamental significance, is our implementation of
the restriction that throat radii are nonnegative.  Here we do this as for
a particle in a half space, leading to the boundary condition~$\psi(0,t)=0$.
Other implementations might be used, the most general condition being only
that the wave function~$\psi$ entail no current in the $-R$~direction at~$R=0$.
Our condition eliminates from consideration any processes such as wormhole
creation or disconnection at~$R=0$.  Including these processes would
drastically alter the physics of the model---essentially, from the quantum
mechanics of one variable to quantum field theory---and would require a
formalism for describing the topology changes.  It should not, however, alter
the instability suggested by our results.  At issue is the stability of the
spacetime foam, of which an individual wormhole is just one fluctuation.
Certainly for any particular wormhole, the probability of growth to large
size is affected strongly by the inclusion or exclusion of topology change.
Indeed, since the time scale implied by our results for the wave function
to leak to large radii is much longer that that for a classical bounce, the
chance that a specific wormhole grows large should be much smaller than that
it pinches off and disappears at zero radius, if that is allowed with more
than an extremely small probability.  But given the possibility of topology
change, the foam should contain an equilibrium population of holes fluctuating
into and out of existence.  If it is possible for a hole to grow large, this
population will eventually give rise to some large holes.  Again the analogy
may be drawn to the $\alpha$~decay of a heavy nucleus:  $\alpha$~particles
continually form and disperse within the nucleus, on a time scale typically
much shorter that that of the decay; the instability represented by the
tunnelling of an $\alpha$~particle out of the nucleus remains.

The more sophisticated analyses needed to probe the quantum dynamics of
spacetime beyond the restrictions and limitations of these calculations
present a considerable challenge.  Our results suggest, however, that
this is an aspect of quantum gravity theory well worth such consideration.\par
\blankline
\centerline{\bf ACKNOWLEDGMENTS}\par\nobreak
We thank M.~Visser, C.~M.~Will, and K.~Young for many useful discussions.
This work was supported by the U.~S.~National Science Foundation via Grants
Nos.~PHY91--16682 and PHY89--22140 at Washington University in St.~Louis, and
No.~PHY91--05935 at the University of Wisconsin--Milwaukee.
\blankline
\hrule
\item{[1]}See, e.g., J.~J.~Halliwell, Int.~J.~Mod.~Phys.~{\bf A5,} 2473 (1990);
in {\it Quantum Cosmology and Baby Universes,} Proceedings of the 7th
Jerusalem Winter School, Jerusalem, Israel, 1990, edited by S.~Coleman,
J.~Hartle, T.~Piran, and S.~Weinberg (World Scientific, Singapore, 1991),
Vol.~7, pp.~159--243.

\item{[2]}S.~W.~Hawking, Phys.~Rev.~D {\bf 37,} 904 (1988); S.~B.~Giddings
and A.~Strominger, Nucl.~Phys.~{\bf B307,} 857 (1988); S.~Coleman,
Nucl.~Phys.~{\bf B307,} 867 (1988); {\bf B310,} 643 (1988); also papers
in Ref.~[1].

\item{[3]}I.~H.~Redmount and W.-M.~Suen, Phys.~Rev.~D {\bf 47,} R2163 (1993).

\item{[4]}J.~A.~Wheeler, Ann.~Phys.~(NY) {\bf 2,} 604 (1957);
{\it Geometrodynamics\/} (Academic, New York, 1962), pp.~71--83.

\item{[5]}R.~P.~Geroch, J.~Math.~Phys.~{\bf 8,} 782 (1967); P.~Yodzis,
Commun.~Math.~Phys.~{\bf 26,} 39 (1972).

\item{[6]}See, e.g., W.~Unruh, Phys.~Rev.~D {\bf 40,} 1053 (1989) and
references cited therein.

\item{[7]}M.~S.~Morris, K.~S.~Thorne and U.~Yurtsever,
Phys.~Rev.~Lett.~{\bf 61,} 1446 (1988); V.~P.~Frolov and I.~D.~Novikov,
Phys.~Rev.~D {\bf 42,} 1057 (1990); J.~Friedman, M.~S.~Morris, I.~D.~Novikov,
F.~Echeverria, G.~Klinkhammer, K.~S.~Thorne, and U.~Yurtsever,
{\it ibid.}~{\bf 42,} 1915 (1990); K.~S.~Thorne, in {\it Nonlinear Problems
in Relativity and Cosmology,} Proceedings of the 6th Florida Workshop in
Nonlinear Astronomy, Gainesville, Florida, 1990, edited by J.~R.~Buchler,
S.~L.~Detweiler, and J.~R.~Ipser (New York Academy of Sciences, New York,
1991), Vol.~631, pp.~182--193; S.~W.~Hawking, in Proceedings of the 6th Marcel
Grossmann Meeting on General Relativity, Kyoto, Japan, 1991, edited by
H.~Sato and T.~Nakamura (World Scientific, Singapore, 1992), Part~A, pp.~3--13.

\item{[8]}T.~A.~Roman, Phys.~Rev.~D {\bf 33,} 3526 (1986); {\bf 37,} 546
(1988).

\item{[9]}M.~Visser, Phys.~Rev.~D {\bf 39,} 3182 (1989); {\bf 41,} 1116
(1990); Nucl.~Phys.~{\bf B328,} 203 (1989).

\item{[10]}W.~Israel, Nuov.~Cim.~{\bf 44B,} 1 (1966); {\bf 48B,} 463~(E)
(1967); C.~W.~Misner, K.~S.~Thorne, and J.~A.~Wheeler, {\it Gravitation\/}
(Freeman, San Francisco, 1973), pp.~551--555; S.~K.~Blau, E.~I.~Guendelman,
and A.~H.~Guth, Phys.~Rev.~D {\bf 35,} 1747 (1987).

\item{[11]}M.~Visser, Phys.~Lett.~{\bf B242,} 24 (1990); Phys.~Rev.~D {\bf 43,}
402 (1991).

\item{[12]}I.~H.~Redmount, W.-M.~Suen, and K.~Young (in preparation).

\item{[13]}T.~Newton and E.~Wigner, Rev.~Mod.~Phys.~{\bf 21,} 400 (1949);
see also S.~S.~Schweber, {\it An Introduction to Relativistic Quantum Field
Theory\/} (Row and Peterson, Evanston, 1961), pp.~54--62.

\item{[14]}I.~H.~Redmount and W.-M.~Suen, Int.~J.~Mod.~Phys.~A {\bf 8,}
1629 (1993).

\item{[15]}J.~B.~Hartle, Phys.~Rev.~D {\bf 38,} 2985 (1988).

\item{[16]}C.~W.~Misner, K.~S.~Thorne, and J.~A.~Wheeler, {\it Gravitation\/}
(Freeman, San Francisco, 1973), frontispiece.

\item{[17]}R.~M.~Wald, {\it General Relativity\/} (University of Chicago Press,
Chicago, 1984), pp.~453--459.

\item{[18]}P.~H\'aj\'\i\v{c}ek, Commun.~Math.~Phys.~{\bf 150,} 545 (1992);
P.~H\'aj\'\i\v{c}ek, B.~S.~Kay, and K.~V.~Kucha\v{r}, Phys.~Rev.~D {\bf 46,}
5439 (1992).

\item{[19]}L.~S.~Schulman, {\it Techniques and Applications of Path
Integration,} (Wiley, New York, 1981), pp.~92--96.

\item{[20]}Schulman~[19], pp.~40--41.

\item{[21]}Schulman~[19], pp.~32--39, 118.

\item{[22]}Schulman~[19], pp.~42--44.\par
\vfill\eject
\centerline{\sl Figure Captions}
\blankline
\blankline
FIG.~1.  Amplitude of the wormhole propagator $G^{\rm (WKB)}[R,t;R_0,0]$,
for $R_0=10$ and $t=20$, all quantities in Planck units.
\blankline
FIG.~2.  Squared magnitude of a wormhole wave function~$\psi(R,t)$
evolving via the propagator~$G^{\rm (WKB)}$.  The initial wave function is a
real Gaussian, as described in the text.  All quantities are in Planck units.
\blankline
FIG.~3.  Magnitude of the dominant contribution~$\psi$ to a wormhole wave
function taken to be initially unity near the radius $R_0=10$ Planck lengths,
evaluated at the same radius after $10^4$--$10^6$ Planck times.
\blankline
FIG.~4.  Behavior of the WKB propagator at large imaginary time~$\tau$, here
in units of the Planck time.  The initial and final radii are fixed at
10~Planck lengths.  Shown are the real and imaginary parts of the
$\tau$ derivative of~$\ln G$.  If the Feynman-Kac ground-state energy
[Eq.~(4.4) of the text] existed, the real part would approach that value
and the imaginary part would tend to zero at large~$\tau$.
\vfil\eject\end